\documentclass{article}

\usepackage{lineno,hyperref}
\usepackage{amsmath,amssymb}
\usepackage{graphicx}

\usepackage[utf8]{inputenc}
\usepackage[english]{babel}
\usepackage[T2A]{fontenc}

\usepackage{comment}

\modulolinenumbers[10]
\modulolinenumbers[1]
\newcommand\nnfootnote[1]{%
  \begin{NoHyper}
  \renewcommand\thefootnote{}\footnote{#1}%
  \addtocounter{footnote}{-1}%
  \end{NoHyper}
}

\begin{document}
\title{Synchronization of wave structures in a heterogeneous multiplex network of 2D vdP lattices with attractive and repulsive intra-layer coupling}
\author{I.A. Shepelev \footnotemark[1], S.S. Muni \footnotemark[2], T.E. Vadivasova\footnotemark[1]}
{\renewcommand\thefootnote{\fnsymbol{footnote}}%
  \footnotetext[1]{Department of Physics, Saratov State University, 83 Astrakhanskaya Street, Saratov, 410012, Russia}\footnotetext[2]{School of Fundamental Sciences, Massey University, Palmerston North, New Zealand}}
\maketitle
\begin{abstract}
We explore numerically the synchronization effects in a heterogeneous two-layer network of two-dimensional (2D) lattices of van der Pol oscillators.The inter-layer coupling of the multiplex network has an attractive character. 
One layer of 2D lattices is characterized by attractive coupling of oscillators and demonstrates a spiral wave regime for both the local and nonlocal interaction. The oscillators in the second layer are coupled through active elements and the interaction between them has the repulsive character. We show that the lattice with the repulsive type of coupling demonstrates complex spatiotemporal cluster structures, which can be called as labyrinth-like structures. 
We show for the first time that this multiplex network with fundamentally different types of intra-layer coupling demonstrates the mutual synchronization and a competition between two types of structures.
Our numerical study indicates that the synchronization threshold and the type of spatiotemporal patterns in both the layers strongly depend on the ratio of the intra-layer coupling strength of the two lattices. 
We also analyze the impact of intra-layer coupling ranges on the synchronization effects.
\end{abstract}

\nnfootnote{Keywords: Synchronization, multiplex networks, spatiotemporal patterns, repulsive coupling, van der Pol oscillator, spiral wave}
\nnfootnote{E-mail addresses: I.A. Shepelev (\url{igor\_sar@li.ru}), S.S. Muni(\url{s.muni@massey.ac.nz}), T.E. Vadivasova (\url{vadivasovate@yandex.ru})}
\newpage
\section*{Extended abstract}
A lot of objects in nature and technical systems represent complex network with ensembles of nonlinear elements interacting between each other through different coupling. They can be  networks  of neurons, populations of living organisms, ensembles of interacting quantum oscillators, computer networks, power grids and many more. In addition, different multicomponent system (the networks and ensembles) can interact between each other and form a complex multilayer network. Exploring such systems with different topology is currently one of the most relevant directions in nonlinear dynamics.

The fundamental phenomenon of synchronization plays an important role in the dynamical behavior of complex multicomponent systems. It leads to partial or complete oscillation coherence of individual elements. The synchronization in spatially distributed ensembles and networks is reflected by the formation of spatiotemporal patterns with different complexity. When  layers (sub-ensembles) of the multilayer network interact, there is not only the synchronization of  the temporal dynamics, but also the partial or complete synchronization of spatiotemporal structures and wave processes. The degree and nature of heterogeneity of interacting layers also plays a very important role.

The heterogeneity is inherent in most real multicomponent systems. Multilayer networks can be used for a simulation of real complex systems, which include different subsystems. In this case, the layers can consist of elements of different types. For example, the layers can correspond to groups of neurons with various characteristics, populations of diverse living organisms, different types of transport coupled in a network, certain energy sources included in a single energy system, etc. But even if all the elements of the network are identical, the layers of the network can significantly differ in the intra-layer coupling. For example, elements of one network layer can be coupled locally or not at all, while elements of another layer are characterized by nonlocal interaction. Such models are often used in neurodynamics. However, it is not only the topology of the links that is important, but also their nature. In a number of works, attractive and repulsive types of coupling are considered. They lead to significant differences in the behavior of networks of oscillators. The repulsive coupling is especially interesting as the neuron interaction in certain cases is repulsive and can be efficient in modeling neuronal networks.. Networks with the attractive coupling demonstrate the effects of partial or complete in-phase synchronization, wave regimes, chimera states. On the contrary, the repulsive coupling impedes the in-phase synchronization and often leads to the effect of oscillation death. In general, the dynamics of the network becomes  simpler. The question arises, if a multilayer network consists of heterogeneous  layers, in which the heterogeneity is associated with the attractive and repulsive character of the intra-layer coupling, then what will be dynamics of the interacting layers and what structures will be predominant?
   
The present work aims to give an answer to this question by using an example of a two-layer network, which consists of two 2D lattices of self-sustained oscillators with attractive and repulsive intra-layer coupling. We consider a possibility of the mutual synchronization of the lattices and its features. We also analyze what  structure will  prevail depending on the ratio of  intra-layer coupling coefficients.

\section{\label{sec:intro} Introduction}

The phenomenon of synchronization is one of the most important dynamical effects in nature. It plays a crucial  role in the collective dynamics of complex multi-component systems and networks ~\cite{PIK01S, NEK02S, MOS02C, OSP07S, ANI14D, BOC18S,mosekilde2002chaotic,balanov2008synchronization}. The emerging field of synchronization has a great importance in the study of applied issues, for example, in neuronal networks~\cite{BAT17M, RAM19P,buzsaki2006rhythms}, social networks~\cite{GIR02C}, transport systems~\cite{CAR13M}, technological networks~\cite{HNG10E, MEN14H, WNG16E,dorfler2012synchronization}, etc. Real networks in general are characterized by a very complex topology. For this reason, numerical simulation of these systems is sufficiently complicated and often impossible. However, the study of simplified models enables us to find out the main features in the network dynamics, in particular the synchronization effects. Recently,  exploring the dynamics of multilayer networks has become one of the actual directions in nonlinear dynamics~\cite{BOC14T, KIV14M, LEE15T, DOM16T}. The latter better describe and simulate the real world systems. For example, neural systems of the brain are characterized by a complex multilayer topology \cite{shipp2007structure,meunier2010modular}, namely neurons interacts through electrical and chemical synaptic connection. This leads to the coexistence of domains with spatially coherent and incoherent behavior \cite{rattenborg2000behavioral,hizanidis2016chimera}.
In multilayer topology, a network consists of several layers (in each layer nodes can interact with each other in different ways) coupled with various types of inter-layer coupling. 

Different types of synchronization have been explored in multiplex networks, namely inter-layer synchronization \cite{GAM15I, RAK17T}, generalized synchronization, external and mutual synchronization ~\cite{STR18SC, RYB19F}, adaptive synchronization \cite{KAS18SC}, explosive synchronization~\cite{LEY13E, ZHN15E, JAL19I} and remote synchronization \cite{SAW18DC,RYB20RC}. There are a number of research works \cite{LEY17I,ghosh2016birth, majhi2017chimera,ghosh2018non} devoted to the synchronization of complex spatiotemporal patterns in multiplex networks with non-identical layers.  
These works have explored the influence of different factors of both intra-layer and inter-layer interactions on the synchronization of structures in both layers of the network. These factors can be time delays in the coupling \cite{ghosh2016birth, majhi2017chimera}, different topology of the interaction \cite{ghosh2018non} and proportion of various types of coupling \cite{jalan2017repulsion} within one of the layer. Indeed, different types of intra-layer coupling can significantly change the dynamics of isolated layers and can also lead to noticeable changes in the synchronization effects. 
Despite of a large number of publications studying the interaction of layers in multiplex networks with different nature of intra-layer coupling, there are still many issues in this direction which are not yet explored. One of these issues is about the interaction between layers with the attractive and repulsive coupling. This case has been considered in \cite{jalan2017repulsion} for two coupled 1D ring of the phase oscillators. It has been shown that the repulsive links in the second layer significantly affects the dynamical behavior of the first ring. In particular, they contribute to the formation of chimera states. This leads to many unanswered interesting questions such as a) the dynamical behavior of similar multiplex networks with complex nonlinear elements (for example, van der Pol oscillator) b) the possibility of mutual synchronization of layers with different types of intra-layer coupling c) is full synchronization possible in this multiplex network and, if possible, which synchronous regime will be the most typical?. It should be noted that the attractive and repulsive interaction leads to essentially different dynamics of the isolated layers. The attractively coupled layer demonstrate complete synchronization and different wave regimes such as traveling waves \cite{nekorkin-2002,neu1997initiation,shepelev2017bifurcations}, spiral and target waves \cite{nekorkin-2002,mikhailov2012foundations,luo2020development} as well as various chimera states \cite{kuramoto2002coexistence,abrams2004chimera,Zakharova-2020}, which represents the spatiotemporal patterns with coexisting coherence and incoherence domains.

The repulsive coupling in turn generally impedes the synchronization and leads to the amplitude or oscillatory death\cite{Ullner-2007,Hens-2013,Hens-2014,Nandan-2014}. It has also been shown that this coupling induces oscillations in various excitable systems \cite{yanagita2005pair}, and even leads to the emergence of traveling waves and a regime of partial synchronization \cite{Tsimring-2005,Nandan-2014}. Interest in the study of systems with repulsive interaction is associated with the use of negative (inhibitory) coupling in the model of neurodynamics \cite{balazsi2001synchronization,wang2011synchronous,rabinovich2006dynamical}. 

Note that in most of the previous studies of multilayer networks, the latter consisted of interacting one-dimensional ensembles of nonlinear systems. It can be quite interesting to study synchronization in a multiplex network which includes two-dimensional (2D) lattices of oscillators. It is well known that 2D ensembles with the attractive coupling can demonstrate wave spatiotemporal regimes, such as spiral and target waves~\cite{nekorkin-2002,mikhailov2012foundations,luo2020development} and spiral and target wave chimeras~\cite{SHM04RSW, MAR10SM, TOT18SW, BKH20SF}. This leads to an unexplored question about the dynamics of a 2D lattice with repulsive coupling.

The purpose of this work is to reveal and study synchronization in a heterogeneous two-layer multiplex network consisting of coupled 2D lattices of van der Pol oscillators. In the first lattice, the intra-layer coupling is repulsive (through an element with the negative differential resistance), while  in  the second lattice elements are  coupled through resistance (attractively).
The inter-layer coupling is bidirectional, pairwise and has the same attractive character as the intra-layer coupling in the second layer.
A regime under study in the 2D lattice with the attractive coupling represents well known spiral waves. The lattice with repulsive coupling demonstrates only standing waves with a complex spatial profiles.
We observe the in-phase synchronization between the layers and competition between the structure. We show that a type of structure in the synchronized lattice strongly depends on the value of the intra-layer coupling strength. Besides, we show that the synchronization is possible only within certain ratios of values of the intra-layer coupling strength in the layers. We also analyze how the nonlocality degree of the intra-layer interaction affects the synchronization features.

\section{The model}

\subsection{Network equations}

\begin{figure}[!t]
\centering
\includegraphics[width=0.7\linewidth]{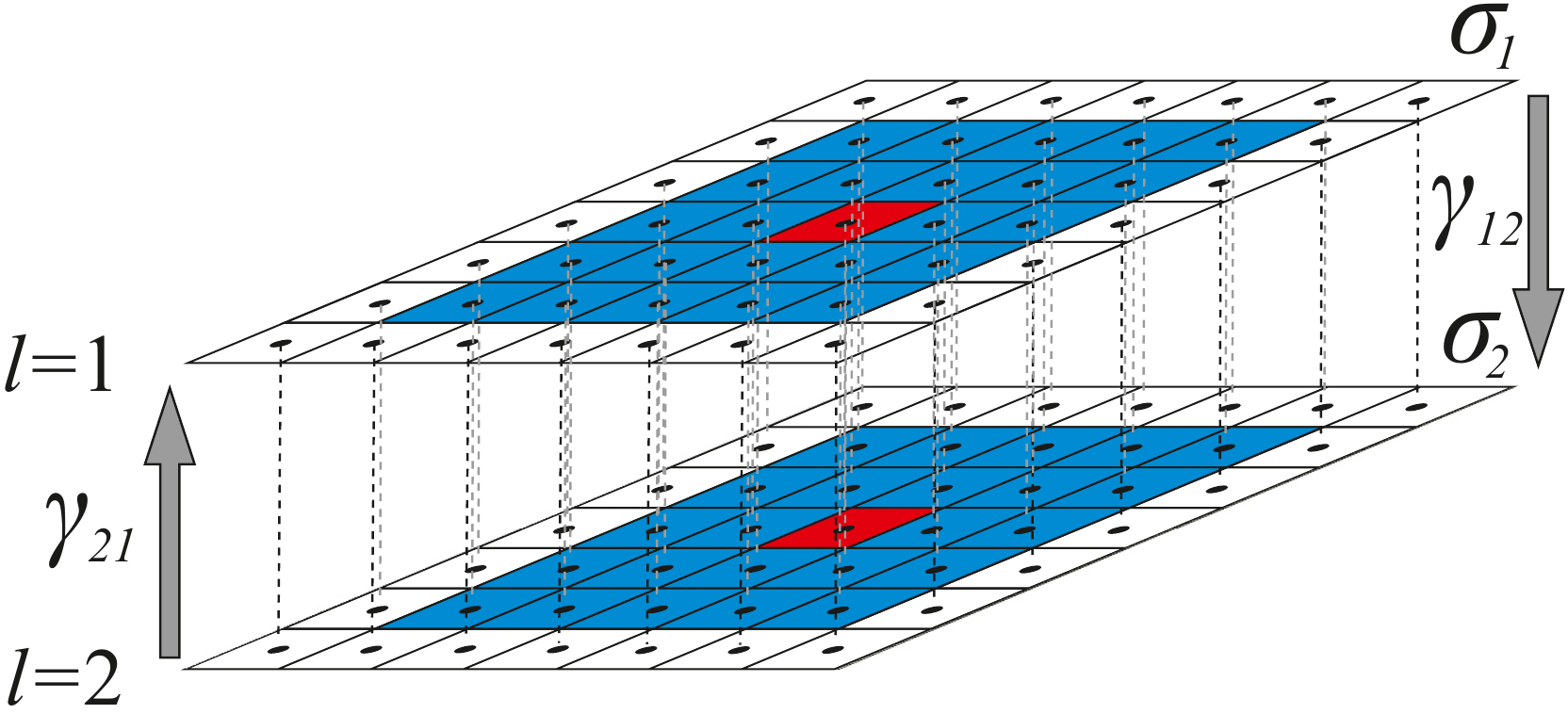}\\
(a)\\[5pt]
\includegraphics[width=0.99\linewidth]{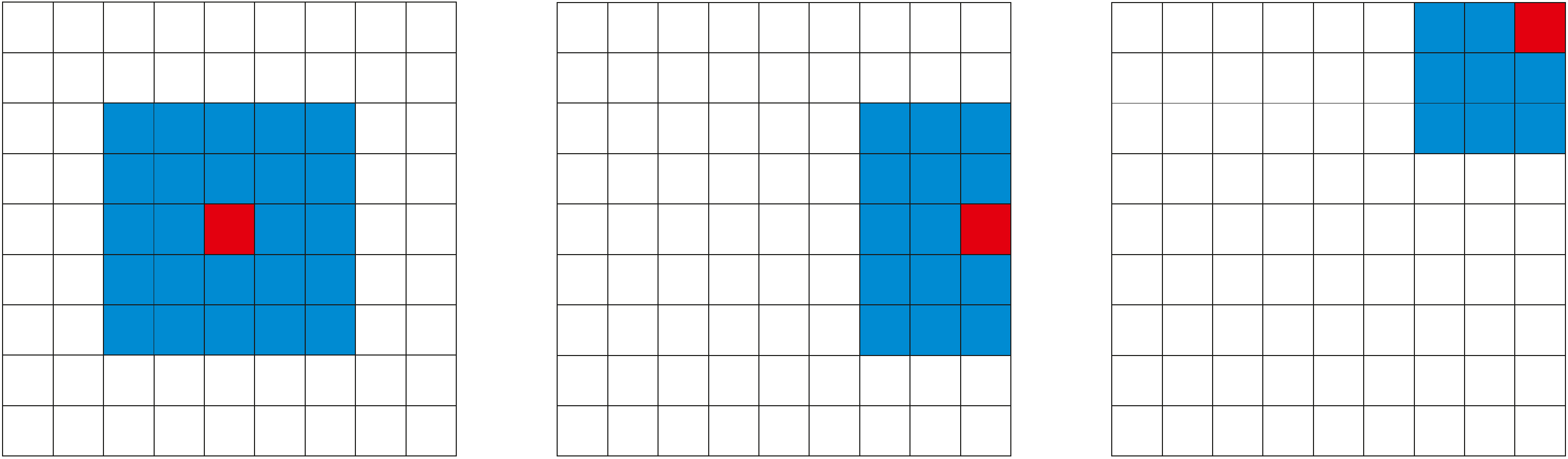}\\
(b)\hspace{75pt}(c)\hspace{75pt}(d)\hspace{0pt}
\caption{(Color online) (a) Scheme of a two-layer multiplex network of coupled 2D lattices which are characterized by the intra-layer coupling strength $\sigma_l$ and the inter-layer coupling $\gamma_{kl}$ ($k,l=1,2$). 
(b)-(d) Schemes of the intra-layer coupling in an isolated 2D lattice for the case of no-flux boundary conditions~Eq.~\eqref{eq:free} for different locations of the selected node (marked in red): (b) in the lattice center, (c) in the middle of the right edge, and (d) in the right upper corner. Oscillators coupled with the selected $(i,j)$th node are depicted in blue, and the rest uncoupled nodes are shown in white. The intra-layer coupling range $P_l=2$.} 
\label{fig_MC_BC}
\end{figure}

We study a multiplex network which consists of two layers and is schematically shown in Fig.~\ref{fig_MC_BC}(a). Each layer represents a two-dimensional (2D) lattice of coupled nonlinear systems and includes $N\times N = 50\times 50$ nodes. The layers are pairwise and bidirectionally coupled with each other. The local dynamics of each element are governed by the van der Pol (vdP) oscillator:
\begin{equation}
\begin{aligned}
\dot{x}=~&y, \\
\dot{y}=~&\varepsilon (1-x^2)y - \omega^2 x,
\end{aligned}
\label{eq:vdP}
\end{equation}
where ($x,\,y $) are dynamical variables, $\varepsilon$ denotes the nonlinearity level, and $\omega$ is the natural frequency of self-sustained oscillations. We fix $\varepsilon =2.0$ and $\omega=2.0$ ensuring the regime of relaxation oscillations in the individual elements.

The multiplex network is described by the following system of equations: 
%
\begin{equation}
\begin{array}{l}
\dot{x}_{i,j}^l = x_{i,j}^l ,
\\[12pt]
\dot{y}_{i,j}^l = y^l_{i,j}+ J_l\dfrac{\sigma_l}{Q_l}
\sum\limits_{m_l,n_l} \left(y_{m_l,n_l}^l - y_{i,j}^l \right)
+\sum\limits_{k=1}^{2} \gamma_{kl} \left(y_{i,j}^k - y_{i,j}^l \right),
\end{array}
\label{eq:grid}
\end{equation}
where ($x_{i,j}^l,y_{i,j}^l$) are dynamical variables, $l$ labels the layer, $l=1,2$, double low indices $(i,j)$, where $i,j = 1, \dots, N=50$, indicates the position of the element on the two-dimensional lattice. The coefficient $\gamma_{kl} $ is the inter-layer coupling strength, while parameter $\sigma_l $ corresponds to the intra-layer coupling strength in the $l $th layer, $l,\,k=1,2$.

The intra-layer coupling is introduced in the second equation in Eq.~\eqref{eq:grid} of each lattice. Parameter $J_l $ determines the type of the intra-layer coupling and is equal to $-1 $ for the first layer and $+1 $ for the second one. This way of the introducing intra-layer links corresponds to repulsive active coupling in the first lattice and to attractive resistive coupling in the second layer.
 $\sigma_l$ is the intra-layer coupling strength, and $P_l$ denotes the sizes of coupling ranges within each layer. 
The parameters $P_l$ determines the number $Q_l$ of all intra-layer links in both directions for each node of the $l$th layer. 
The quantity $Q_l$ for each node of the $l$th layer represents a combination of all the links  with the indices $m_l$ and $n_l$ according to the following relations corresponding to the no-flux boundary conditions:   
\begin{equation}\label{eq:free}\left\{\begin{aligned}
&\max(1,i-P_l) \leqslant m_l \leqslant \min(N,i+P_l), \\
&\max(1,j-P_l) \leqslant n_l \leqslant \min(N,j+P_l),
\end{aligned}\right.\end{equation}
This type of the boundary conditions is illustrated  in Fig.~\ref{fig_MC_BC}(b)-(d) for an isolated lattice and for different locations of the  selected oscillator when the intra-layer coupling range is equal to 2. 
In particular, accordingly to Fig.\ref{fig_MC_BC}(b) a central element is coupled with $Q_l=(2P_l+1)^2-1$ neighbors.

The pairwise bidirectional coupling between the layers is introduced in the second equation of each layer and described by the third term in the system~Eq.~\eqref{eq:grid}. $\gamma_{kl}$ is the inter-layer coupling strength and is defined as follows:
\begin{equation}\label{eq:matr-gamma}
\gamma_{kl}=
\begin{pmatrix}
  0 & \gamma_{12}\\
  \gamma_{21} & 0 
  \end{pmatrix}
\end{equation}
where the row and column numbers are related to the corresponding layers. In our numerical simulation $\gamma_{12}=\gamma_{21}=\gamma$. The coupling range of the inter-layer interaction corresponds to the common multiplex network, namely each $(i,j)$th oscillator of the first layer is linked only with $(i,j)$th element of the second layer.

The initial conditions for all the dynamical variables of the network~Eq.~\eqref{eq:grid} are chosen to be random and uniformly distributed within the interval~$[-1,1]$.

The model equations~\eqref{eq:grid} are integrated using the Runge-Kutta 4th order method with the time step $dt=0.005$. The transient time $T_{tr} $ is equal to $T_{tr}=10000 $ time units for all the cases under study.

\subsection{Dynamics of isolated lattices}

We explore the dynamical regimes which can be obtained in the 2D lattices of coupled vdP oscillators when the interaction between the two layers in the network Eq.~\eqref{eq:grid} is absent ($\gamma=0 $). At first, we consider the dynamics of the lattice with attractive coupling (second layer, $J_2=+1 $ in Eq.~\eqref{eq:grid}). This coupling corresponds to the case when elements interact through a resistance. Three dynamical regimes are observed with the variation of the intra-layer coupling parameters $\sigma_2 $ and $P_2 $. When the coupling strength $\sigma_2 $ is very weak the oscillators behave incoherently with each other. An increase in $\sigma_2 $ leads to the formation of a wave structure, namely the spiral waves. 
\begin{figure}[!b]
\centering
\hspace{-2mm}\parbox[c]{.34\linewidth}{ 
  \includegraphics[width=\linewidth]{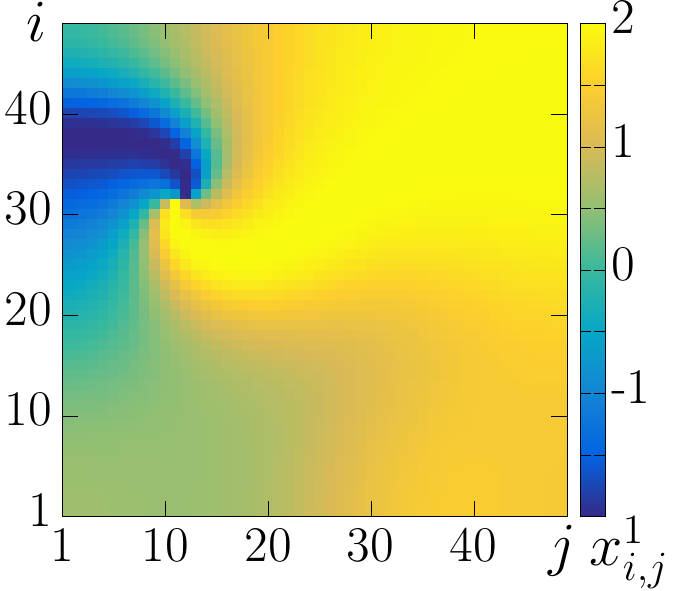}
    \vspace{-7.5mm} \center (a)
}
\hspace{-2mm}\parbox[c]{.34\linewidth}{
\includegraphics[width=\linewidth]{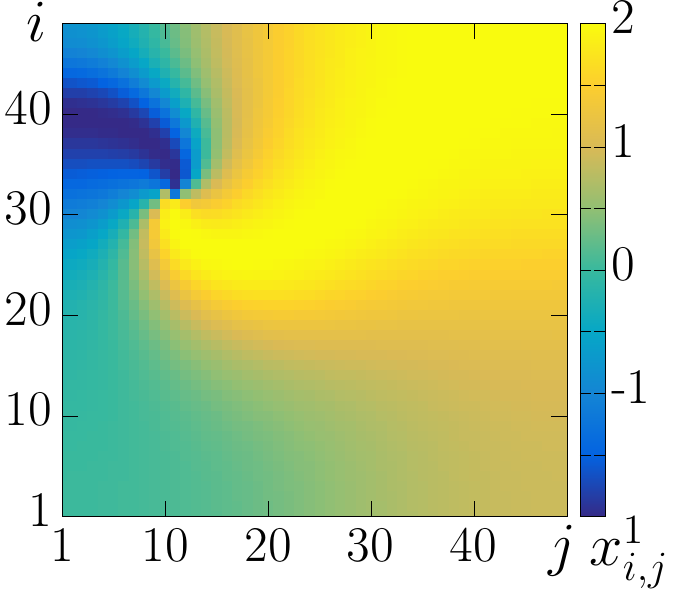}
\vspace{-7.5mm}\center (b)
}
\hspace{-2mm}\parbox[c]{.34\linewidth}{
\includegraphics[width=\linewidth]{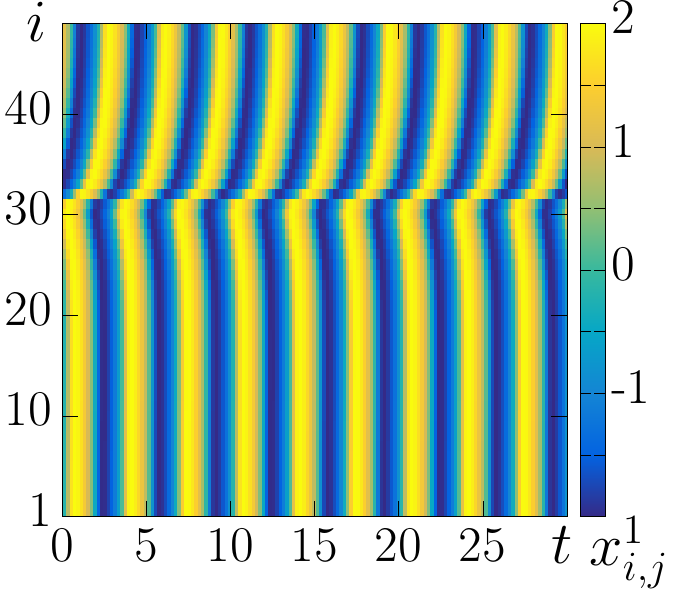}
\vspace{-7.5mm}\center (c)
}
\caption{(Color online) Spiral waves in the second isolated layer (lattice) with attractive coupling of the network Eq.~\eqref{eq:grid} at $\sigma_2=0.1 $. Snapshots of the system state (a) $P_2=1 $ (local coupling) and (b) $P_2=2 $ (nonlocal coupling). (c) is a space-time plot of the cross-section $j=12 $  for the local coupling. Parameters: $J_2=1 $, $\varepsilon=2 $, $\omega=2$, $N=50$.} 
\label{fig:attractive_layer}
\end{figure}
A different number of spiral waves can coexist in the lattice for various set of initial conditions. We study the case when  there is only one spiral wave in the lattice. The spiral waves are observed only for the local $P_2=1 $ and short nonlocal $P_2=2 $ coupling. It should be noted that for the local interaction, the wave structure is formed within significantly longer range of the coupling strength than for the case of nonlocal coupling $P_2=2 $. The third regime is the partial or complete synchronization of element oscillations. It is considered when $P_2 \geq 3$ and   for the case of strong coupling strength and $P_2=1,2 $ (the spiral waves disappear). In the present work we study the lattice in the regime of spiral waves. They are exemplified in Fig.~\ref{fig:attractive_layer}(a) for $P_2=1 $ and in Fig.~\ref{fig:attractive_layer}(b) for $P_2=2 $.
The spatiotemporal dynamics is illustrated in a space time plot of the $j=12 $ cross-section passing through the wave center in Fig.~\ref{fig:attractive_layer}(c). It is seen that the wave front rotates around the wave center. Note, that this spiral wave cannot be transformed into a spiral wave chimera since the wave is destroyed when the coupling range becomes greater than $P_2\geq  3$.

\begin{figure}[!h]
\centering
\parbox[c]{.4\linewidth}{
\vspace{-3.5mm}\center 
  \includegraphics[width=\linewidth]{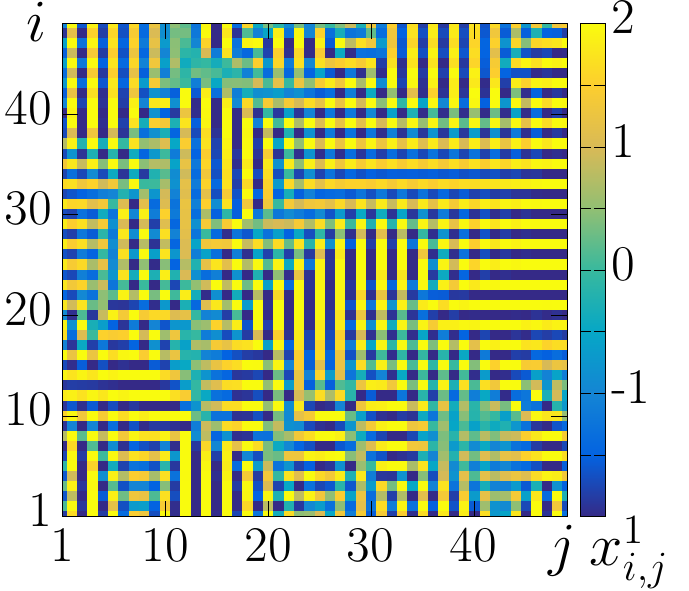}
    \vspace{-8.5mm} \center (a)
\includegraphics[width=\linewidth]{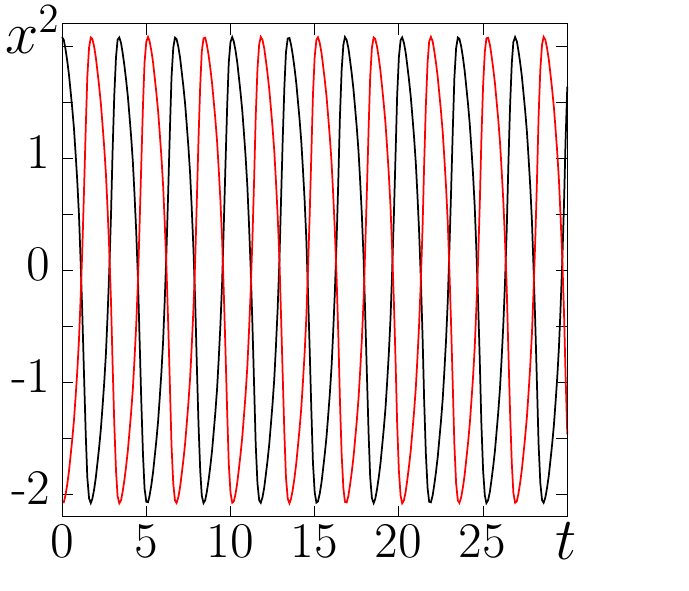}
\vspace{-8.5mm}\center (c)
}
\parbox[c]{.4\linewidth}{
\vspace{-3.5mm}\center   
  \includegraphics[width=\linewidth]{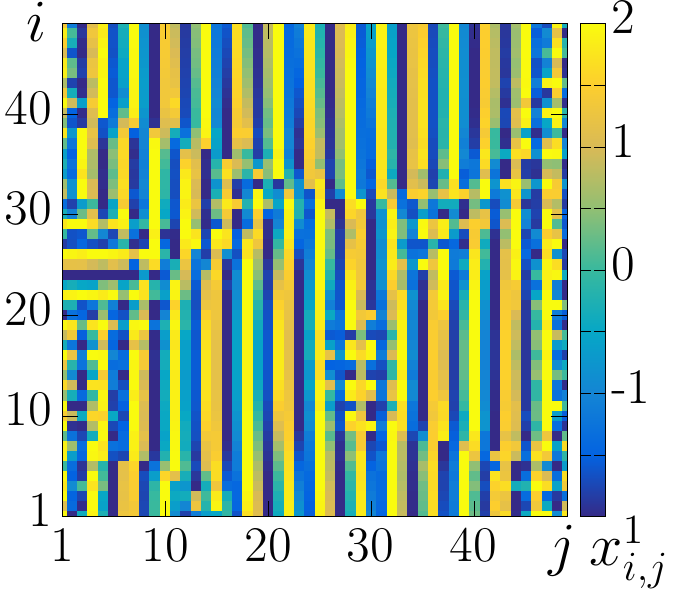}
   \vspace{-8.5mm} \center (b)
\includegraphics[width=\linewidth]{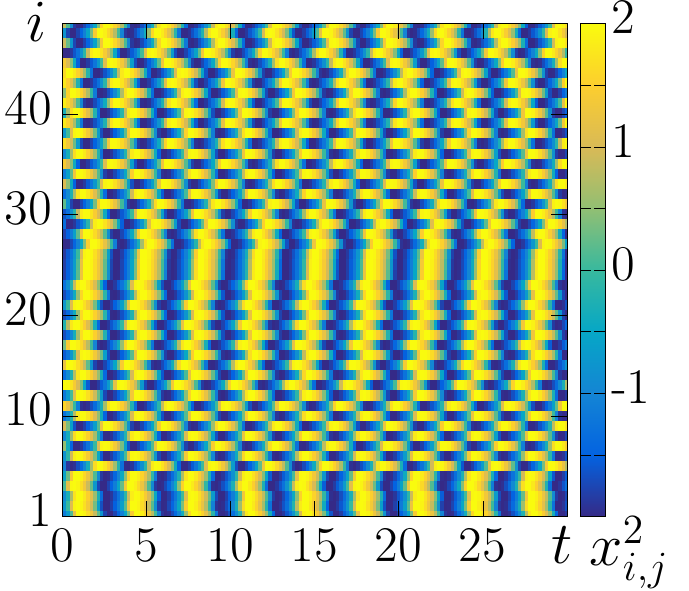}
 \vspace{-8.5mm} \center (d)
}
\caption{(Color online) Labyrinth-like structures in the first isolated layer (lattice) with repulsive coupling of the network Eq.~\eqref{eq:grid} for $\sigma_1=0.1 $. Snapshots of the system state (a) $P_1=1 $ (local coupling) and (b) $P_1=2 $ (nonlocal coupling). (c) time realizations for the $(10,11) $th (black line) and $(11,10) $th oscillators, (d) a space-time plot of the cross-section $j=30 $  for the local coupling. Parameters: $J_1=-1 $, $\varepsilon=2 $, $\omega=2$, $N=50$.}
\label{fig:repuslsive_layer}
\end{figure}

The repulsive type of coupling between vdP oscillators presents an interaction through an active element with the linear negative differential resistance ($J_1=-1 $ in Eq.\ref{eq:grid}). Our study shows that this coupling leads to the appearance of a completely different spatiotemporal regime in the lattice and that the spiral waves cannot be obtained for any initial conditions. Examples of the typical regimes for the local ($P_1=1 $) and nonlocal ($P_1=2 $) coupling are exemplified in Fig.~\ref{fig:repuslsive_layer}(a) and (b), respectively.
The spatiotemporal structure presents alternation of  ``strips'', in each of which all  elements oscillate with the very similar phase. However, in the two adjacent strips the  oscillation phases are shifted by the half period, i.e. the oscillations are in anti-phase. These strips have different length and can be located in either vertical or horizontal direction. Similar spatiotemporal patterns have been discovered in \cite{SHEPELEV2021105513} in the lattice of strongly coupled vdP oscillators. These structures have been called ``labyrinth-like structures''. Despite of the qualitative similarity between the state in these two systems, there are significant changes between them. At first, in \cite{SHEPELEV2021105513} a reason behind the formation of such structures is the emergence of two coexisting chaotic attractors in the phase space of individual oscillators induced by the strong coupling. In contrast, for system under study, there is no bistability and oscillations in the neighboring strips differ by only the instantaneous phases. This feature is illustrated in a plot of the time series for two selected oscillators from adjacent strips  in Fig.~\ref{fig:repuslsive_layer}(c). The spatiotemporal dynamics is depicted by a space-time plot of the $j=30 $th cross-section in Fig.~\ref{fig:repuslsive_layer}(d). 
The elongation of the coupling range $P_1 $ leads to a simplification of the spatiotemporal structure (see Fig.~\ref{fig:repuslsive_layer}(b) for $P_1=2 $). The strips with similar phases become wider and the number of strip intersections decreases. For the long coupling ranges the structures become regular and represent an alternation of strips or squares of oscillators with  certain values of the instantaneous phases. At the same time, the quantitative features  of the regime for both short and long coupling ranges remain very similar. This enables us to assume that the same regime with different spatial topology is observed within a wide interval of the coupling range values.
Since the spiral waves in the second lattice exist only for the coupling range $P_2=1 $ and   $P_2=2 $, we will study structures in the first lattice for the same values of $P_1 $.

It should be noted that  all the structures under study presented above in Figures~\ref{fig:attractive_layer} and \ref{fig:repuslsive_layer} are sufficiently robust and are not  qualitatively and quantitatively changed when the intra-layer coupling strength are varied within the interval $\sigma_{1,2}\in [0.05,0.15] $.
 For this reason we will study the inter-layer interaction $\gamma$ for the cases when value of the intra-layer coupling strength $\sigma_{1,2} $ is varied within the mentioned interval. 

The structures presented above are chosen to be the initial states in the first and second lattices. Besides, using these initial states we have obtained a set of initial conditions for each values of the intra-layer coupling strength $\sigma_{1,2}\in [0.05,0.15] $ and use them in studying the synchronization effects in the multiplex network in ~Eq.~\eqref{eq:grid}. 

\subsection{Synchronization measure}

To quantify and distinguish synchronization effects between the layers in the network~Eq.~\eqref{eq:grid}, we calculate the correlation coefficient between the corresponding ($i,j$)th pairs of oscillators of the first and second layers  as below:
\begin{equation} \label{eq-corr}
 R^{12}_{i,j} = \dfrac
{\overline{ \tilde x_{i,j}^1 \cdot \tilde x_{i,j}^2}}
{\sqrt{\overline {{ (\tilde x_{i,j}^1)}^2} 
\cdot \overline  {{(\tilde x_{i,j}^k)}^2} }}, \quad \tilde x_{i,j}^{12} = x_{i,j}^{12} - \overline{ x_{i,j}^{12}} ,
\end{equation}
where $\overline \cdots $ means time-averaging over $T_{\rm av}=10 000$ time units.
Using the correlation measure  Eq.~\eqref{eq-corr} for arbitrary oscillation modes, one can evaluate effective synchronization between the layers when $R_{i,j}^{12}\approx \pm1$ but $R_{i,j}^{12}\neq \pm1$ and complete synchronization when $R_{i,j}^{12}= \pm 1$.  Besides, the correlation coefficient value enables one to diagnose the in-phase and anti-phase synchronization of spatiotemporal structures in the regime of periodic oscillations of the lattice nodes. It is $+1$  in the first case and $-1$ in the second one. 
Note that since our network~Eq.~\eqref{eq:grid} is heterogeneous (the coupling in different layers has different character), the synchronization measure in~Eq.~\eqref{eq-corr} cannot be strictly equal to $\pm 1$ and synchronization effects must be understood in their effective sense. In this case a synchronization criterion must be imposed. Synchronization is assumed to take place if $|R_{i,j}^{12}|\geq 0.95$. It should be noted that this condition is arbitrary and any value of $R_{i,j}$ close to $\pm 1$ may be chosen as a threshold. 

\section{Mutual synchronization when varying the coupling strength in the  layers}

We now introduce the bidirectional inter-layer coupling $\gamma$ between the layers to the $y$ dynamical variable in the multiplex network~Eq.~\eqref{eq:grid}  and explore synchronization effects between them. The inter-layer coupling has the same attractive (dissipative) character as the intra-layer coupling in the second layer. 
To highlight the influence of the nonlocality of intra-layer coupling on the synchronization effects we study two cases of mutual synchronization, namely when the intra-layer coupling is local $P_{1,2}=1 $ and nonlocal $P_{1,2}=2 $.

\subsection{Synchronization in the network of locally coupled lattices}

We now study the synchronization effects for the case of an interaction of the two 2D lattices with the  local intra-layer coupling. These lattices are in the regimes exemplified in Fig.~\ref{fig:repuslsive_layer}(a) for the first layer and in Fig.~\ref{fig:attractive_layer}(a) for the second lattice. Our study shows that the synchronization feature of these structures strongly depends on values of the intra-layer coupling in the lattices. Besides, in-phase synchronization is observed within a wide range of the inter-layer coupling strength $\gamma $. To describe and illustrate these effects in detail we fix one of the intra-layer coupling strength $\sigma_{1,2}=0.1$ and vary the second one within a range $\sigma_{2,1}\in[0.05,0.15]$ as well as the $\gamma\in [0,1] $ with sample steps $\Delta \sigma_{2,1}=0.01 $ and $\Delta \gamma=0.01 $.
 
\begin{figure}[!ht]
\centering
\parbox[c]{.49\linewidth}{
\vspace{-3.5mm}\center 
  \includegraphics[width=\linewidth]{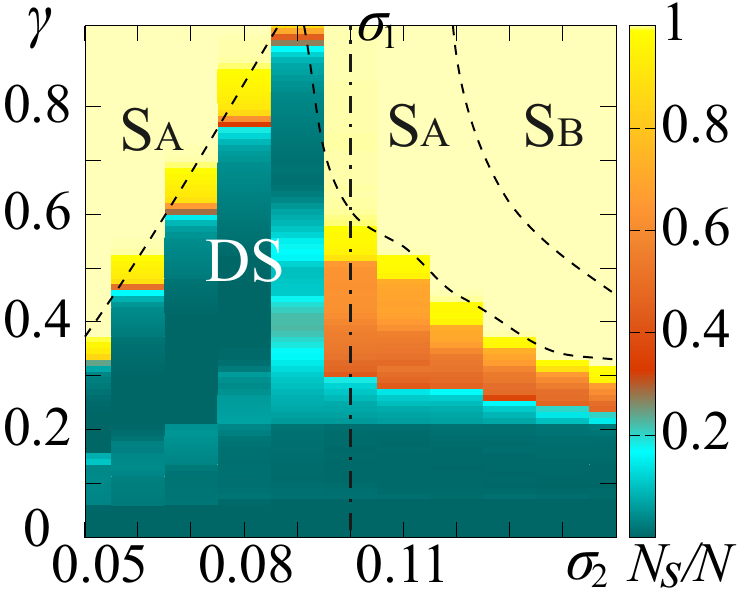}
    \vspace{-7.5mm} \center (a)
}
\parbox[c]{.49\linewidth}{
\vspace{-3.5mm}\center   
  \includegraphics[width=\linewidth]{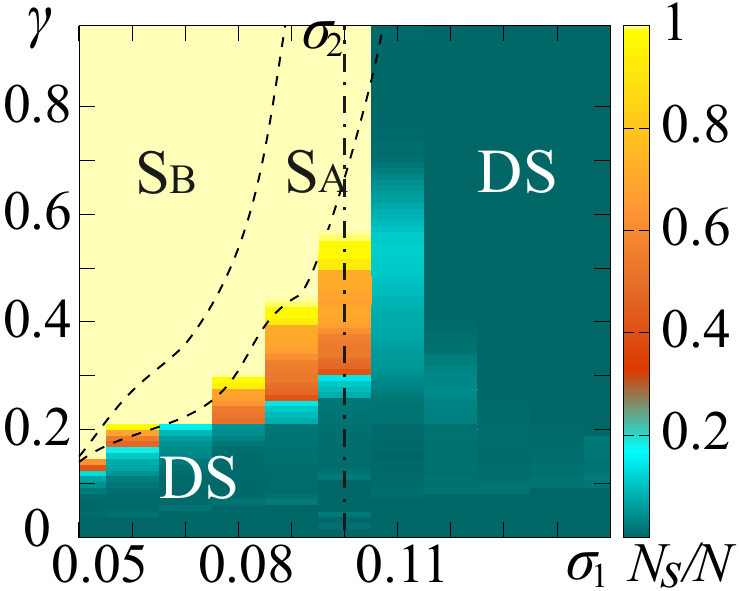}
   \vspace{-7.5mm} \center (b)
}
\caption{(Color online) 2D diagrams of a ratio $N_s/N $ of a number of synchronized $i,j $th oscillator pairs $N_s $ to whole number of pairs $N $ and  synchronous and desynchronized regimes of the network Eq.~\eqref{eq:grid} with local intra-layer coupling $R_{1,2}=1 $ in (a) ($\sigma_2, \gamma$) parameter plane at $\sigma_1=0.1 $, and in (b) ($\sigma_1, \gamma$) parameter plane at $\sigma_2=0.1 $. Region DS corresponds to an absence of the structure synchronization, while in regions $\rm{S_A} $ and $\rm{S_B} $ the lattices are synchronized. In region $\rm{S_A} $ the same labyrinth-like structures (initial state of the first layer) are formed in both lattice, and in region $\rm{S_B} $ the identical spiral waves (initial regime of the second lattice) are observed in the layers. Boundaries between the regions are illustrated by dotted lines. Parameters: $J_1=-1 $, $J_2=1 $, $\varepsilon=2 $, $\omega=2$, $N=50$.}
\label{fig:regime_diagram_local}
\end{figure}

Synchronization effects are detected by evaluating and analyzing the correlation coefficient $R^{12}_{i,j}$. To diagnose synchronization between the layers, we plot two 2D diagrams of synchronous and desynchronized regimes in the ($\sigma_2, \gamma$) parameter plane (for fixed $\sigma_1=0.1 $, Fig.~\ref{fig:regime_diagram_local}(a)) and in the ($\sigma_1, \gamma$) plane (for fixed $\sigma_2=0.1 $, (Fig.~\ref{fig:regime_diagram_local}(b)) by using the values of $\langle R^{12}\rangle $, where $\langle \cdots \rangle $ means that values of the correlation coefficients are averaged over all the oscillator pairs ($i,j$). The color scheme in the diagrams corresponds to the ratio of a number of synchronized ($\langle R^{12}\rangle \geq 0.95$) oscillator pairs $N_s $ to a whole number of elements in the each lattice $N $.
In order to get better insight into the effects of the synchronization in the network~Eq.~\eqref{eq:grid}, we highlight the following region with different dynamical regimes, namely region of desynchronization (denoted by DS), where the spatiotemporal behavior in the two lattices are not identical ($\frac{N_s}{N}<1 $), and the two regions of synchronization (denoted by $\rm{S_A}$ and $\rm{S_B}$). Within region  $\rm{S_A} $, a spatiotemporal structure which is similar to the initial state in the first lattice (see Fig.~\ref{fig:repuslsive_layer}(a)) is observed in both the layers, while the synchronized spiral waves are formed in both of the layers in region $\rm{S_B} $. Boundaries of these regimes are illustrated by dotted lines. These plots show that the synchronization effects have a strong and complex dependence on the ratio of $\sigma_1 $ to  $\sigma_2 $. It is seen that the closeness of these values impedes the synchronization. The synchronization is possible only when either $\sigma_2\geq \sigma_1 $ (see Fig.~\ref{fig:regime_diagram_local}(a)) or $\sigma_1\gg \sigma_2$ (Fig.~\ref{fig:regime_diagram_local}(b)). 
\begin{figure}[!t]
\centering
\hspace{-4mm}\parbox[c]{.05\linewidth}{\
\rotatebox{90}{ $\rm{S_B},~\gamma=0.7 $ ~~~~~~  $\rm{S_A},~\gamma=0.5 $ ~~~~~~~   DS, $\gamma=0.3 $}
 }
\hspace{-1mm}\parbox[c]{.325\linewidth}{
\center first layer
  \includegraphics[width=\linewidth]{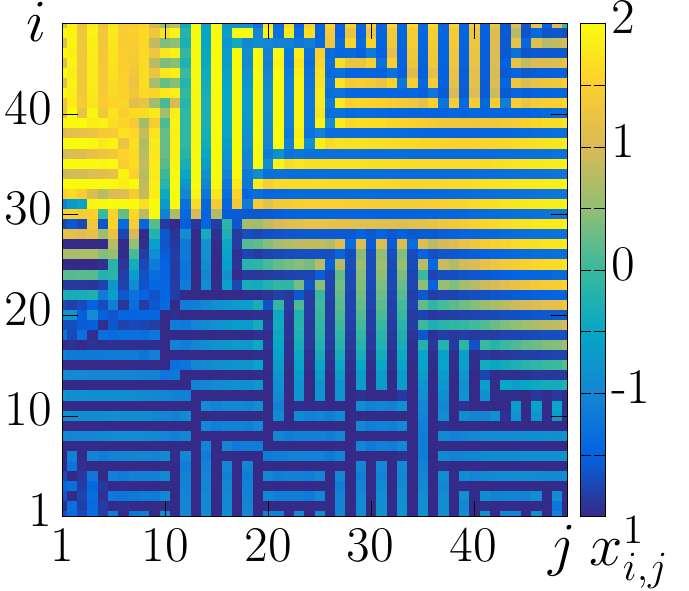}
    \vspace{-8.5mm} \center (a)  
  \includegraphics[width=\linewidth]{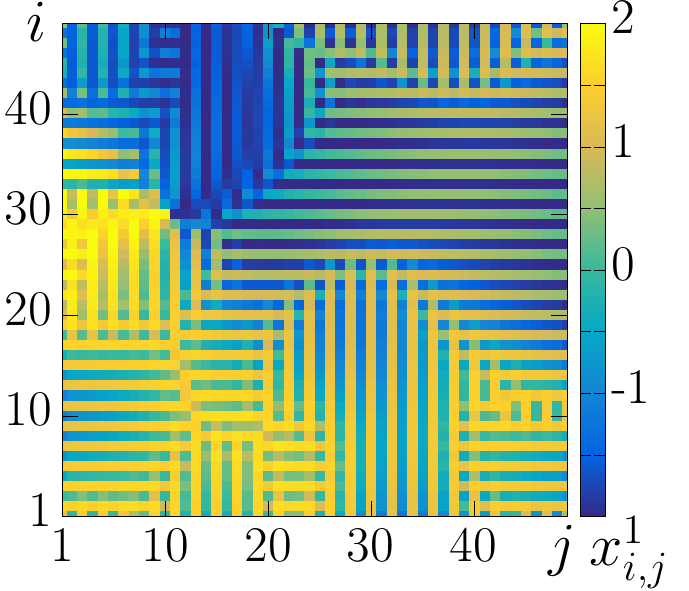}
    \vspace{-8.5mm} \center (d)
   \includegraphics[width=\linewidth]{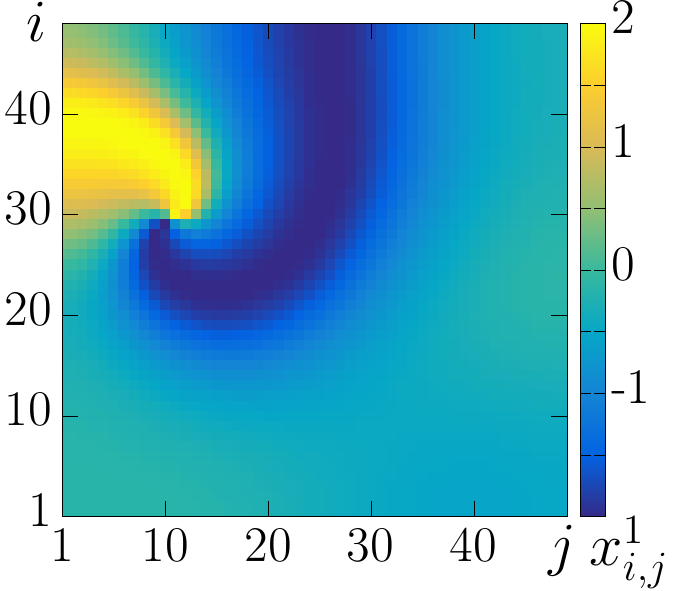}
    \vspace{-8.5mm} \center (g)
}
\hspace{-2mm}\parbox[c]{.325\linewidth}{
\center second layer
\includegraphics[width=\linewidth]{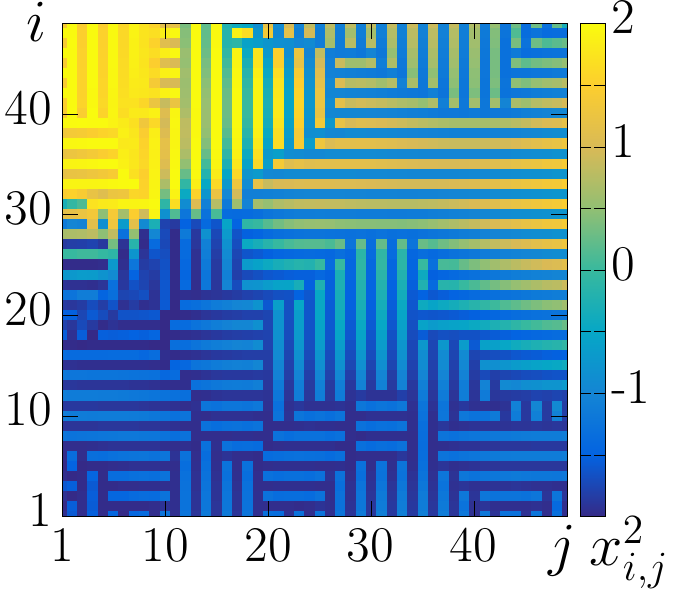}
\vspace{-8.5mm}\center (b)
  \includegraphics[width=\linewidth]{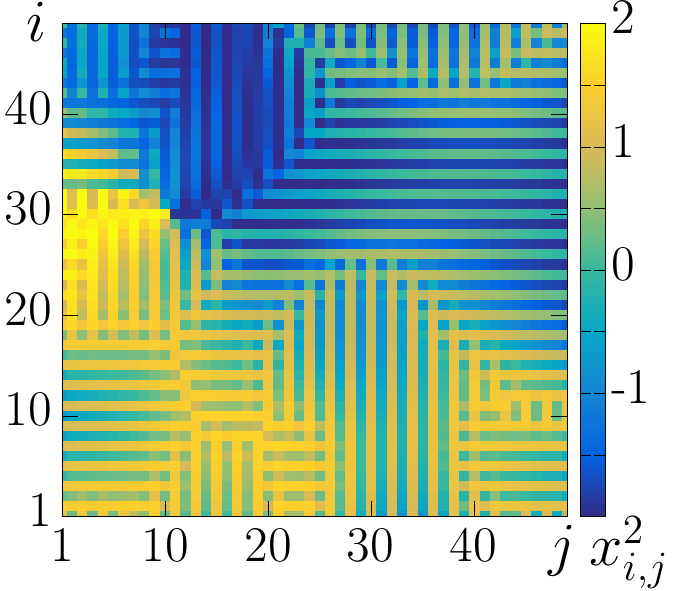}
    \vspace{-8.5mm} \center (e)
  \includegraphics[width=\linewidth]{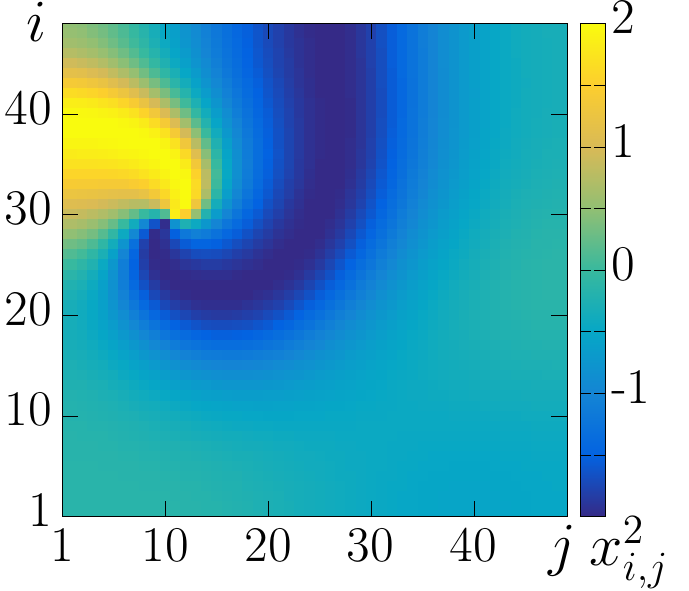}
    \vspace{-8.5mm} \center (h)
}
\hspace{-2mm}\parbox[c]{.325\linewidth}{
\vspace{-.5mm}\center $R^{12}_{i,j} $
\includegraphics[width=\linewidth]{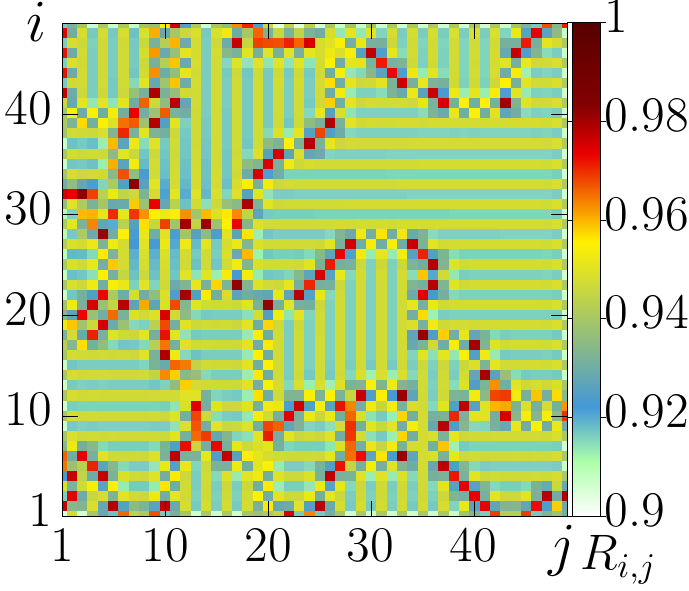}
\vspace{-8.5mm}\center (c)
  \includegraphics[width=\linewidth]{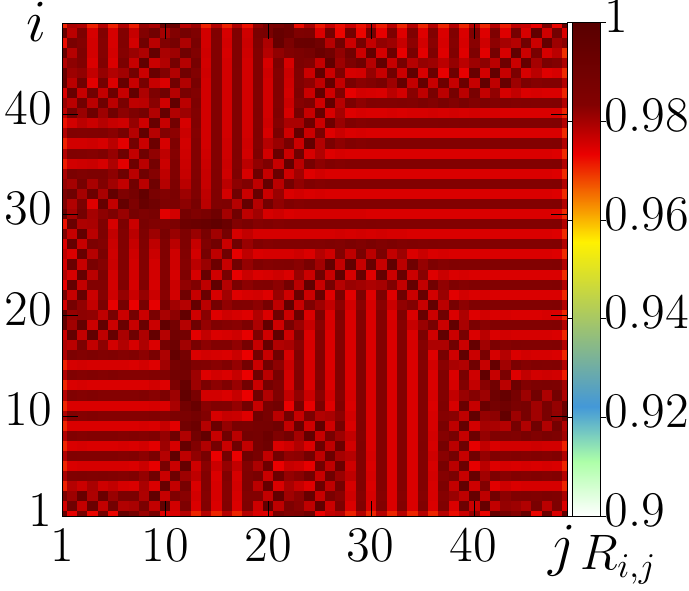}
    \vspace{-8.5mm} \center (f)
  \includegraphics[width=\linewidth]{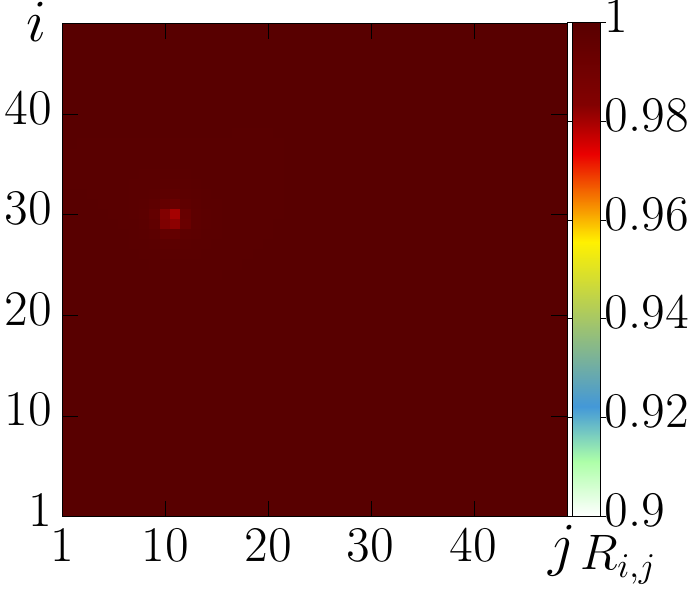}
    \vspace{-8.5mm} \center (i)
}
\caption{(Color online) Typical spatiotemporal structures in the region of diagrams represented in Fig.~\ref{fig:regime_diagram_local} at fixed values of $\sigma_1=0.1 $ and $\sigma_2=0.13 $. Plots in the left and middle columns demonstrate snapshots of the system state for the first and second lattices, respectively. Spatial distribution of the correlation coefficients $R_{i,j} $ are represented in the right column. The top line of plots corresponds to region $DS $ at $\gamma=0.3 $, the median line is for region $\rm{S_A}$  at $\gamma=0.5 $, and the bottom line shows plots for for region $\rm{S_B}$  at $\gamma=0.7 $. Parameters: $J_1=-1 $, $J_2=1 $, $R_{1,2}=1 $, $\varepsilon=2 $, $\omega=2$, $N=50$.} 
\label{fig:sync_P=1}
\end{figure}
Moreover, these results show that the labyrinth-like structure (initial state in the first lattice with the repulsive interaction) is more stable than the spiral wave regime. The first one suppresses the spiral wave within the whole region DS as well as within the synchronous region for the case when values of the inter-layer coupling strength $\gamma $ are sufficiently small and $\sigma_1>\sigma_2 $ (region $\rm{S_A} $). Examples of  structures in the first and second lattices are represented in Fig.~\ref{fig:sync_P=1}(a) and (b) for region DS and in  Fig.~\ref{fig:sync_P=1}(d) and (e) for region $\rm{S_A} $. At first sight, it might seem that the structures in the layers are identical for the both regimes. However, spatial distributions of the correlation coefficient $R_{i,j}^{12} $ illustrated in Fig.~\ref{fig:sync_P=1}(c) (region DS) and  Fig.~\ref{fig:sync_P=1}(f) are noticeably different from each other. For region DS, only a small part of oscillator pairs demonstrate the synchronous behavior (their $R_{i,j}^{12}\geq0.95 $, while rest of the pairs have $R_{i,j}^{12}<0.95 $). This means that the instantaneous phases and/or amplitudes are not the same, i.e. the phase synchronization is absent. The values of correlation coefficients for region $\rm{S_A} $ are $R_{i,j}^{12}\geq0.95 $ for all the $i,j $th oscillator pairs. Hence, the effective in-phase synchronization of the whole system takes place. 
Note that in both the structures there is a certain rotation center around which the wave front propagates and the spatial structure corresponds to the labyrinth-like one.

Significant changes occurs in region $\rm{S_B}$ shown in Fig.~\ref{fig:regime_diagram_local}. The labyrinth-like structure completely disappears and the spiral wave regime is formed in both layers after a certain transient process. This is illustrated by snapshots of the system state for both of the lattices in Fig.~\ref{fig:sync_P=1}(g) and (h). In this case the oscillations are most synchronous in comparison with the other regimes. It is well seen in a spatial distribution of the correlation coefficients $R_{i,j} $ in Fig.~\ref{fig:sync_P=1}(i). Most of the $i,j $th oscillator pairs oscillate in-phase and are characterized by $R_{i,j}\cong 1 $ and only the oscillators around the wave centers have values of $R_{i,j}\lessapprox 1 $. 

When studying the synchronization of locally coupled 2D lattices it has been discovered that the spiral wave regime in the attractively coupled lattice is less stable than the labyrinth-like structure in the lattice with repulsive coupling. The spiral wave can suppress the state in the first lattice only when the intra-layer coupling strength $\sigma_2 $ is stronger than one in the first layer, i.e. $\sigma_2>\sigma_1 $. In other cases, the labyrinth-like structure suppresses the spiral wave.
A question arises how does the non-locality of the intra-layer coupling affects the synchronization effects?

\subsection{Influence of the intra-layer non-local coupling on the synchronization}

We now study the case when values of the intra-layer coupling ranges are 
\begin{figure}[!h]
\centering
\parbox[c]{.49\linewidth}{
\vspace{-3.5mm}\center 
  \includegraphics[width=\linewidth]{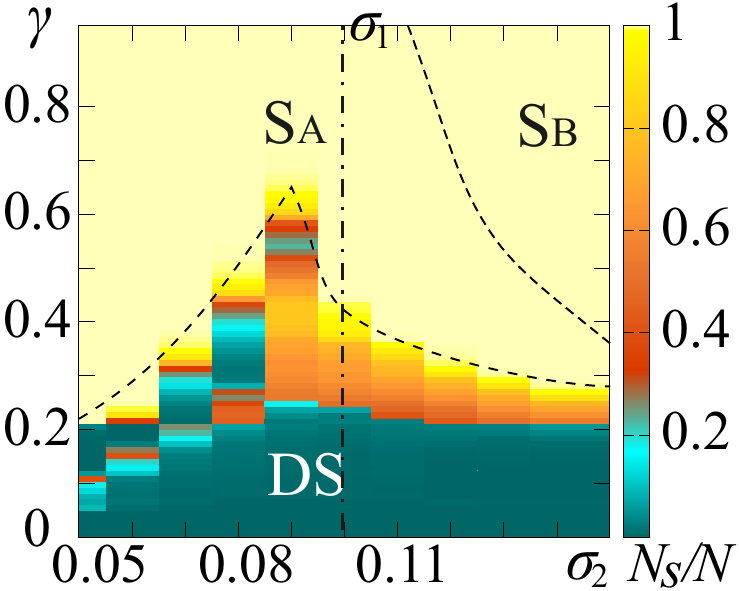}
    \vspace{-7.5mm} \center (a)
}
\parbox[c]{.49\linewidth}{
\vspace{-3.5mm}\center   
  \includegraphics[width=\linewidth]{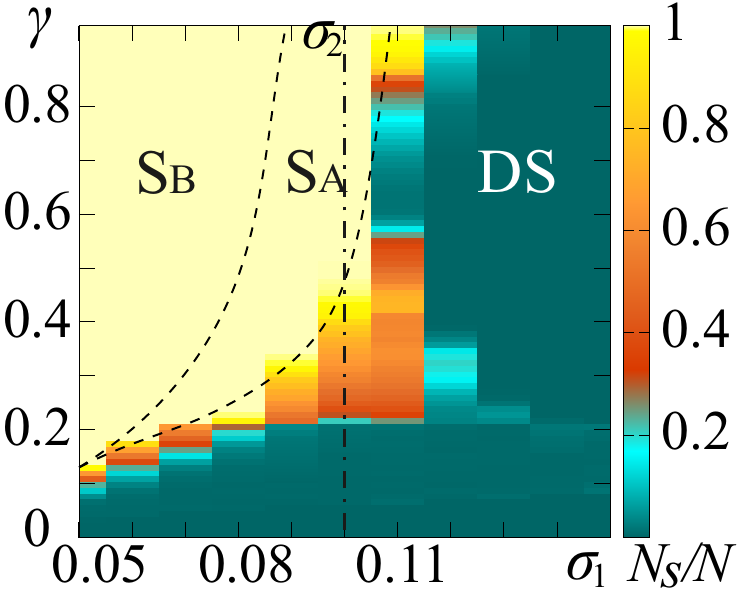}
   \vspace{-7.5mm} \center (b)
}
\caption{(Color online) 2D diagrams of a ratio $N_s/N $ of a number of synchronized $i,j $th oscillator pairs $N_s $ to a whole number of pairs $N $ and  synchronous and desynchronized regimes of the network Eq.~\eqref{eq:grid} with nonlocal intra-layer coupling $R_{1,2}=2 $ in (a) ($\sigma_2, \gamma$) parameter plane at $\sigma_1=0.1 $, and in (b) ($\sigma_1, \gamma$) parameter plane at $\sigma_2=0.1 $. Region DS corresponds to an absence of the structure synchronization, while in regions $\rm{S_A} $ and $\rm{S_B} $ the lattices are synchronized. In region $\rm{S_A} $ the same labyrinth-like structures (initial state of the first layer) are formed in both of the lattice, and in region $\rm{S_B} $ the identical spiral waves (initial regime of the second lattice) are observed in the layers. Boundaries between the regions are illustrated by dotted lines. Parameters: $J_1=-1 $, $J_2=1 $, $\varepsilon=2 $, $\omega=2$, $N=50$.}
\label{fig:regime_diagram_nonlocal}
\end{figure}
equal to $P_{1,2}=2 $. The initial states for these values of $P_{1,2}$ are depicted in Fig.~\ref{fig:repuslsive_layer}(b) for the first repulsively coupled lattice and in Fig.~\ref{fig:attractive_layer}(b) for the second lattice with the attractive coupling. Note that we consider only the case of $P_{1,2}=2$ as the spiral wave exists only for $P_{2}=1 $ and 2.

To diagnose the synchronization between the layers we calculate and plot the 2D diagrams of synchronous and desynchronous regimes in analogy with the diagrams in Fig.~\ref{fig:regime_diagram_local}. They are shown in Fig.~\ref{fig:regime_diagram_nonlocal}.

In general, the non-locality of intra-layer coupling does not lead to qualitative changes in the synchronization features. They have the same dynamical regimes as of the previous case. The inter-layer synchronization is observed for a significantly wider interval of the inter-layer coupling strength  values $\gamma $, especially when values of the intra-layer coupling strength $\sigma_2 $ is less than $\sigma_1 $. Moreover, region $\rm{S_B} $ extended within values of  both $\gamma $ and $\sigma_2 $ (see Fig.~\ref{fig:regime_diagram_nonlocal}(a)). 

\begin{figure}[!h]
\centering
\hspace{-4mm}\parbox[c]{.05\linewidth}{\
\rotatebox{90}{ $\rm{S_B},~\gamma=0.6 $ ~~~~~~  $\rm{S_A},~\gamma=0.4 $ ~~~~~~  DS, $\gamma=0.27 $}
 }
\hspace{-0mm}\parbox[c]{.325\linewidth}{
\center first layer
  \includegraphics[width=\linewidth]{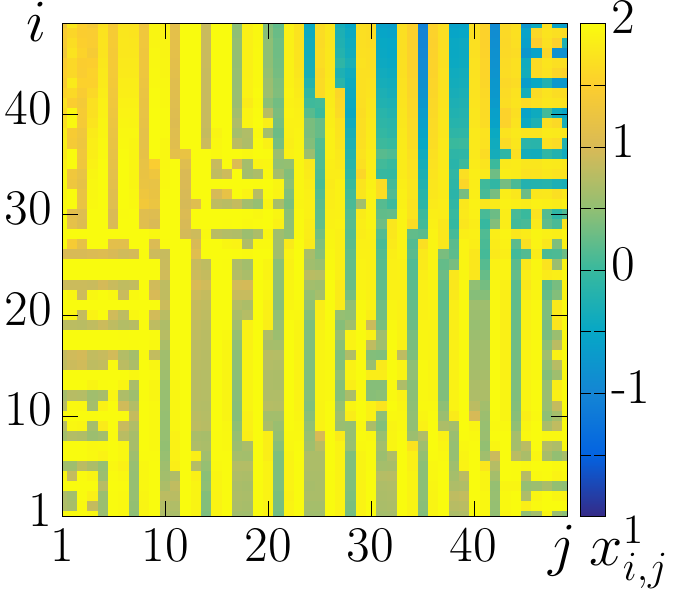}
    \vspace{-8.5mm} \center (a)  
  \includegraphics[width=\linewidth]{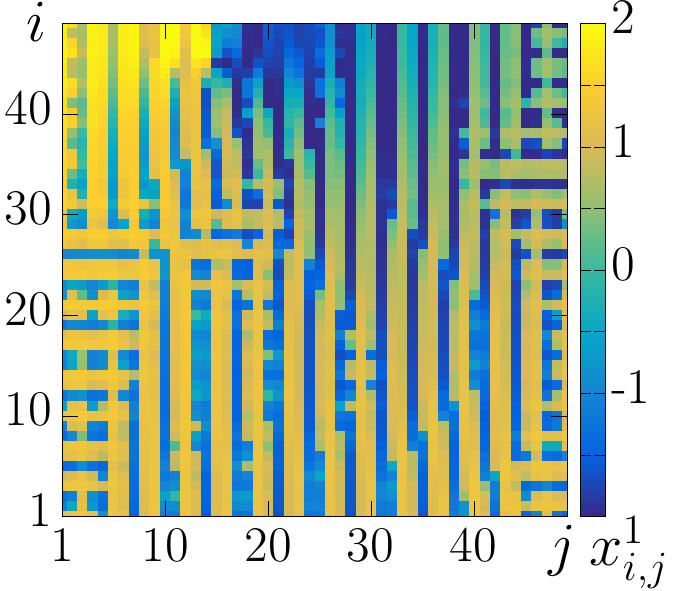}
    \vspace{-8.5mm} \center (d)
   \includegraphics[width=\linewidth]{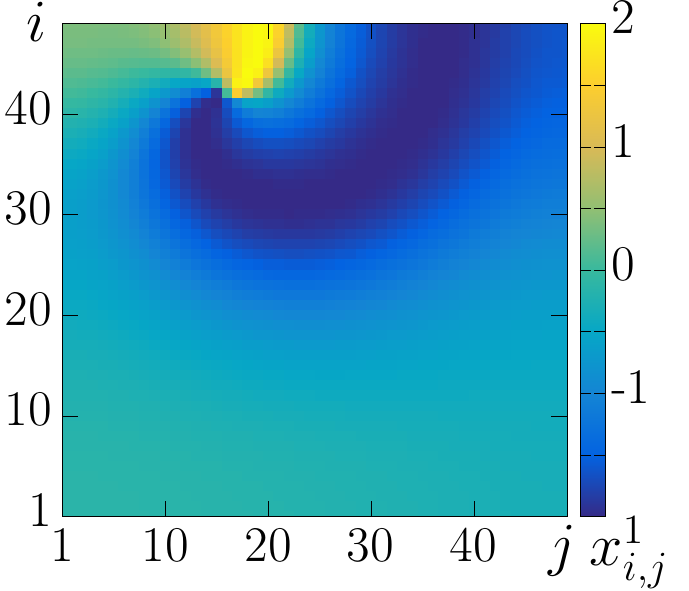}
    \vspace{-8.5mm} \center (g)
}
\hspace{-2mm}\parbox[c]{.325\linewidth}{
\center second layer
\includegraphics[width=\linewidth]{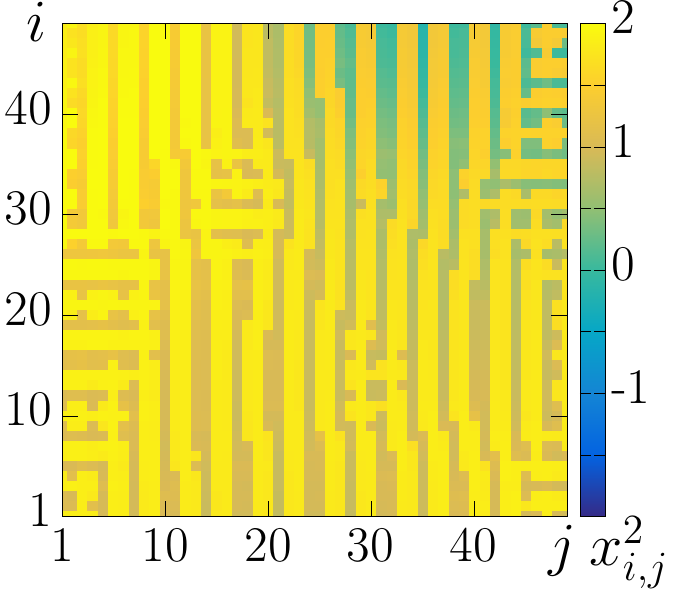}
\vspace{-8.5mm}\center (b)
  \includegraphics[width=\linewidth]{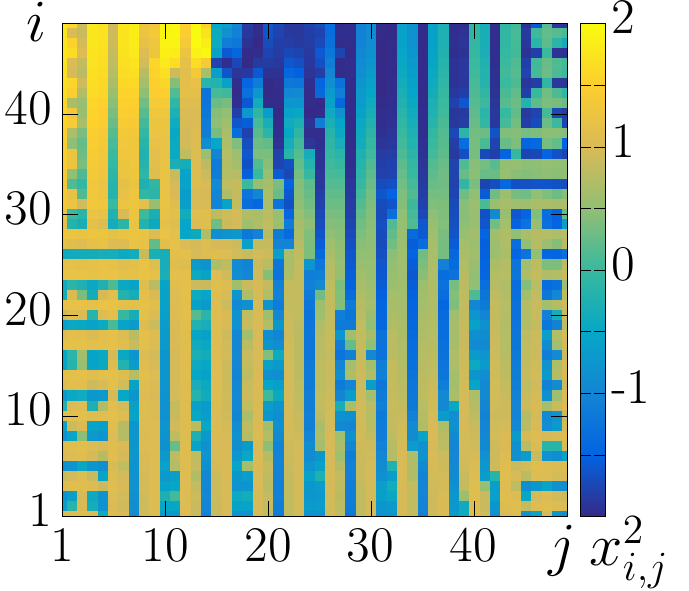}
    \vspace{-8.5mm} \center (e)
  \includegraphics[width=\linewidth]{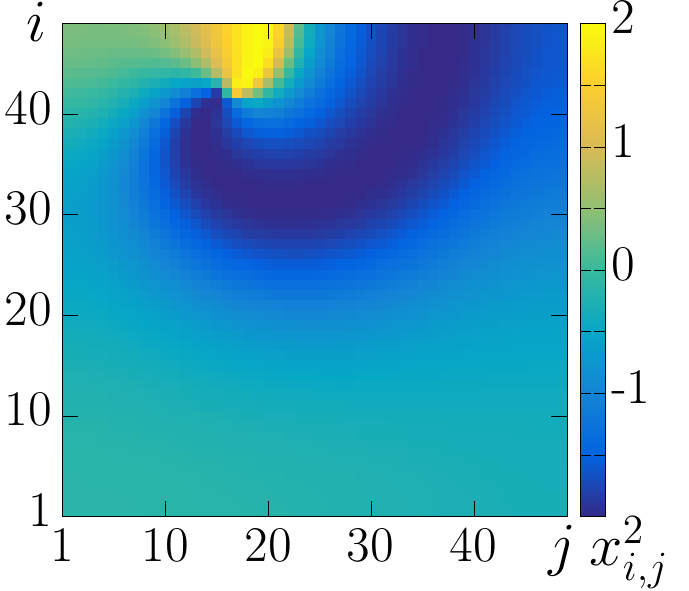}
    \vspace{-8.5mm} \center (h)
}
\hspace{-2mm}\parbox[c]{.325\linewidth}{
\vspace{-.5mm}\center $R^{12}_{i,j} $
\includegraphics[width=\linewidth]{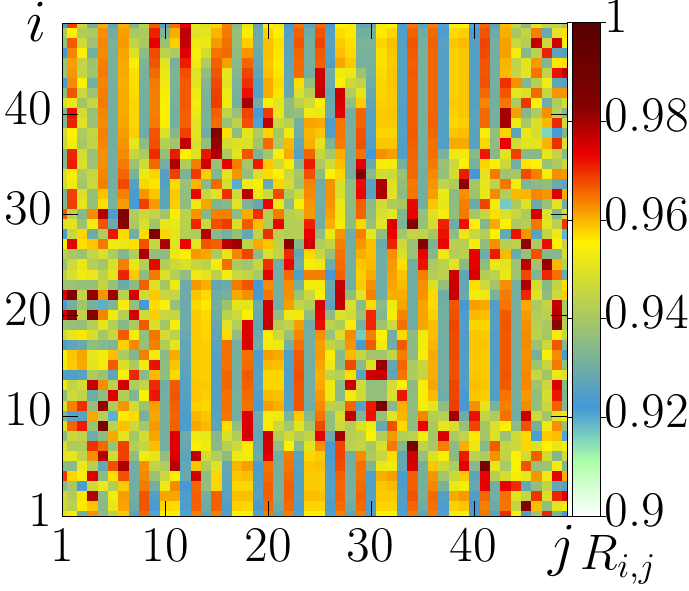}
\vspace{-8.5mm}\center (c)
  \includegraphics[width=\linewidth]{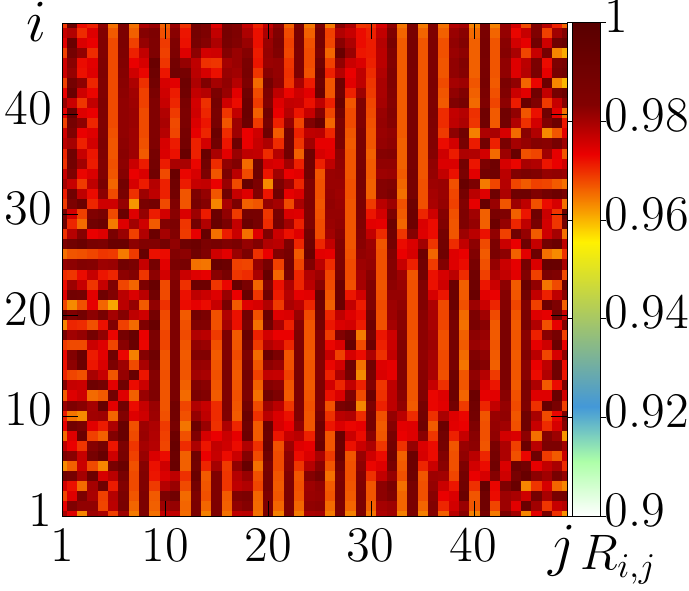}
    \vspace{-8.5mm} \center (f)
  \includegraphics[width=\linewidth]{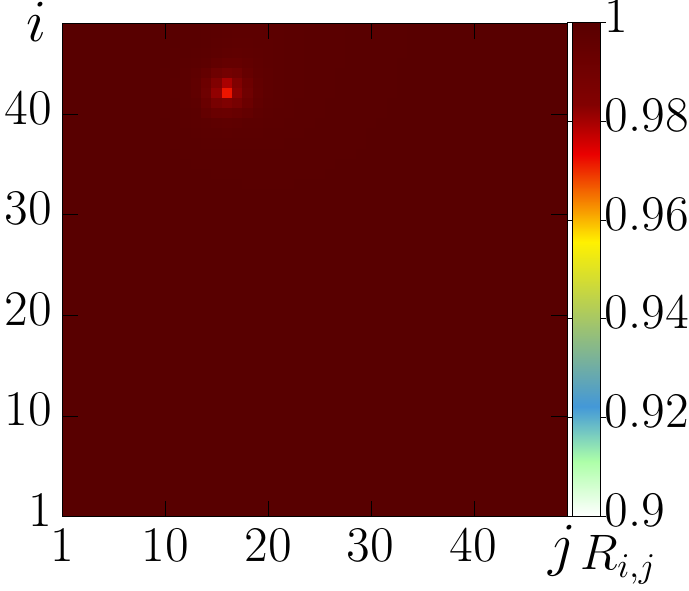}
    \vspace{-8.5mm} \center (i)
}
\caption{(Color online) Typical spatiotemporal structures corresponding to different regions in the diagram of Fig.~\ref{fig:regime_diagram_nonlocal} for fixed values of $\sigma_1=0.1 $ and $\sigma_2=0.13 $. Plots in the left and middle columns demonstrate snapshots of the system state for the first and second lattices, respectively. Spatial distribution of the correlation coefficients $R_{i,j} $ are represented in the right column. The top line of plots corresponds to region $DS $ at $\gamma=0.27 $, the median line is for region $\rm{S_A}$  at $\gamma=0.4 $, and the bottom line shows plots for for region $\rm{S_B}$  at $\gamma=0.6 $. Parameters: $J_1=-1 $, $J_2=1 $, $R_{1,2}=2 $, $\varepsilon=2 $, $\omega=2$, $N=50$.} 
\label{fig:sync_P=2}
\end{figure}
Examples of the structures for each region in the diagrams of regimes in Fig.~\ref{fig:regime_diagram_nonlocal} are illustrated   by snapshots of the system states for the first and second layers in Fig.~\ref{fig:sync_P=2} (the left and middle columns, respectively).  Corresponding spatial distributions of the correlation coefficients are shown in Fig.~\ref{fig:sync_P=2} (the right column). It is seen that the synchronization effects for the case of $P_{1,2}=2 $ are very similar to the case of $P_{1,2}=1 $, despite the fact that the initial  spatiotemporal structures (for $\gamma=0 $) in the layers are noticeably different than in the case of local intra-layer coupling $P_{1,2}=1 $.

\section{Conclusion}

We have explored numerically the synchronization effects in a  multiplex network consisting of pairwise and bidirectionally coupled 2D lattices of van der Pol oscillators. The intra-layer interaction between elements in the first lattice has the repulsive character, while the interaction in the second lattice is attractive and dissipative. Both type of coupling have simple physical interpretation. Our study has shown that these type of intra-layer coupling can lead to a significantly different dynamics in the isolated layers. The repulsively coupled lattice demonstrate the so-called labyrinth-like structures \cite{SHEPELEV2021105513} within a wide range of the coupling parameters. A regime typical in the second layer is that of a spiral wave. It is obtained only for the cases of local and short nonlocal coupling.  These spatiotemporal structures have been chosen as initial states in both of the layers. Note that for any initial combinations, spiral waves are never observed in the lattice with the repulsive coupling, while the labyrinth-like structures are absent in the attractively coupled layer.

It has been shown for the first time that the multiplex network of coupled 2D lattices with fundamentally different types of the intra-layer coupling can demonstrate mutual synchronization of spatiotemporal structures when the attractive and dissipative (resistive) inter-layer coupling $\gamma $ is introduced between the layers. The calculations have revealed that a ratio of the intra-layer coupling strengths $\sigma_{1}/\sigma_{2} $ plays a very important role in the synchronization features, namely the threshold level of intra-layer coupling strength when the effective synchronization occurs and a type of structures in the interacting layers. A regime in the synchronized layers always corresponds to only one of the initial regime, and not a combination of these structures. Thus, a competition between the initial regimes takes place at the mutual interaction between layers.  When the effective synchronization between the layers is absent, the spiral wave in the second layer is destroyed and the same labyrinth-like structures are formed in both of the lattices for any values of the intra-layer and inter-layer coupling strengths. However, the instantaneous phases of oscillations of corresponding elements in the lattices are slightly shifted against each other. If a value of the coupling strength in the first layer is greater than one in the second lattice then the same labyrinth-like structures are formed in both of the lattices. When  the coupling strength in the first layer is greater, the synchronization becomes impossible. On the other hand, if the coupling in the second layer is greater than one in the second layer then labyrinth-like regime in the first lattice can be fully replaced with the spiral wave. In this case, the oscillations in the layers are the most synchronous. Thus, the introduction of bidirectional inter-layer coupling leads to in-phase synchronization and to the formation of the structures which cannot be observed in the isolated lattices, namely the spiral wave in the layer with the repulsive coupling and the labyrinth-like structure in the layer with the attractive coupling.

It has been shown that in the case of nonlocal coupling in both layers, the qualitative features of synchronization remains very similar to the case of local interaction. However, the synchronization of the spatiotemporal structures begins for lower values of the inter-layer coupling strength. Furthermore, the region in which the synchronized spiral waves are observed in both of the lattices expanded.

In our work we have used the correlation coefficient~Eq.~\eqref{eq-corr} as a synchronization measure. This measure enables to diagnose phase relations, namely in-phase or anti-phase synchronization of spatiotemporal structures. Unfortunately, we cannot compare our findings with the data of previously published works. Note that  other synchronization criterias used in \cite{GAM15I, LEY17I, RAK17T} did not allow judging of the phase relations.

\section*{Acknowlegements}
The reported study has been funded by the Russian Science Foundation (project No.~20-12-00119). S.S.M acknowledges the use of New Zealand eScience Infrastructure (NeSI) high performance computing facilities as part of this research.

\bibliography{ArxivMultRepHet}
\bibliographystyle{iopart-num}

\end{document}